\documentclass[preprint2]{aastex}
\usepackage{spr-astr-addons}
\usepackage{graphicx,natbib}
\newcommand{\msol}{M_\odot}

\begin{document}
\title{Gap Formation in the Dust Layer of 3D Protoplanetary Disks}

\shorttitle{Planetary Gaps in 3D Disks}        
\shortauthors{Maddison et al.}

\author{S. T. Maddison}
\affil{Centre for Astrophysics and Supercomputing, Swinburne University, PO Box 218, Hawthorn, VIC 3122, Australia}
\email{smaddison@swin.edu.au}
\author{L. Fouchet\altaffilmark{1} and J.-F. Gonzalez}
\affil{Universit\'e Lyon 1, CRAL (CNRS-UMR 5574), \'Ecole Normale Sup\'erieure de Lyon, 46 all\'ee d'Italie,
F-69364 Lyon cedex 07, France}
\altaffiltext{1}{ETH Z\"urich, Schafmattstrasse 16, HPF D19, CH-8093 Z\"urich, Switzerland}

\begin{abstract}
We numerically model the evolution of dust in a protoplanetary disk using a
two-phase (gas $+$ dust) Smoothed Particle Hydrodynamics (SPH) code, which 
is non-self-gravitating and locally isothermal.  The code follows the three 
dimensional distribution of dust in a protoplanetary disk as it interacts 
with the gas via aerodynamic drag.  In this work, we present the evolution 
of a disk comprising 1\% dust by mass in the presence of an embedded 
planet for two different disk configurations: a small, minimum mass solar 
nebular (MMSN) disk and a larger, more massive Classical T Tauri star (CTTS) 
disk.  We then vary the grain size and planetary mass to see how they effect 
the resulting disk structure.

We find that gap formation is much more rapid and striking in the dust layer 
than in the gaseous disk and that a system with a given stellar, disk and 
planetary mass will have a different appearance depending on the grain size 
and that such differences will be detectable in the millimetre domain with
ALMA.  For low mass planets in our MMSN models, a gap can open in the dust 
disk while not in the gas disk.  We also note that dust accumulates at the 
external edge of the planetary gap and speculate that the presence of a planet 
in the disk may facilitate the growth of planetesimals in this high 
density region.
\end{abstract}


\section{Introduction}
\label{sect-intro}

The effect of a planet in a gaseous disk has been well studied both
analytically and numerically  
\citep{PapLin1984,Kley1999,Bryden-etal1999,dValBorro-etal2006}.
Tidal torques resulting from the gravitational perturbation of the planet 
lead to an exchange in angular momentum which creates a gap around the 
planet.
To sustain the gap in viscous disks, there needs to be a balance between 
the tidal torques, which clear the gap, and viscous torques, which fills the gap
\citep{LinPap1979}. Thus the gap criterion is given by
\begin{equation}
 \frac{M_{\rm p}}{M_{\star}}  >   40 \alpha_{\rm SS} \left(\frac{H}{r_{\rm p}}\right)^2 \, ,
\label{eqn-gapinvisc}
\end{equation}
where $M_{\rm p}$ and $M_{\star}$ are the mass of the planet and star, $\alpha_{\rm SS}$ is the 
\citet{ShakuraSunyaev1973} viscosity parameter, $H$ 
is the disk scale height and $r_{\rm p}$ is the semi-major axis of the planet.
Our previous simulations  \citep[][hereafter BF05]{BF05} show  the 
settling rate -- and hence the thickness of the dust layer -- depends on grain 
size.  Since the gap criterion depends on the disk scale height, this would suggest 
that it is easier to create and sustain a gap in the dust layer than in the 
gas.

There are a variety of observational signature of planetary gaps, including
mid-infrared dips \citep[e.g.][]{Calvet-etal2002,Rice-etal2003} 
and direct scattered light 
\citep{Weinberger-etal1999,Schneider-etal2006} 
and sub-millimetre observations of protoplanetary disks 
\citep{Ozernoy-etal2000,Wilner-etal2002}. 
%

Recent models have indicated that ALMA will be able to detect planetary gaps
at sub-millimetre wavelengths to distances of 100~pc
\citep{Wolf05,Varniere-etal2006}.  However, these models assume that the gas
and dust are well-mixed within the disk and yet we know the dust-to-gas ratio 
changes substantially as grains settle to the mid-plane and migrate radially (BF05).
As well as the gas-to-dust ratio varying throughout the disk, we expect that
the effects of planetary gaps will be stronger in the dust phase than in the 
gas phase, which will further affect observations. 

In this paper we study the formation of a gap triggered by an embedded planet 
in the dust layer of a protoplanetary disk. We will study the effects of planet
mass and grain size on gap formation and evolution in 3D dusty-gas
protoplanetary disks.


\section{Code description and simulation parameters}

We use our 3D, two-phase (gas$+$dust), locally isothermal, 
non-self-gravitating code based on the Smoothed Particles Hydrodynamics (SPH) 
algorithm.
The dusty gas is approximated by two inter-penetrating flows that interact by 
aerodynamic drag. For the nebula parameters used in
this study, we are in the Epstein drag regime and hence 
\begin{equation}
F_\mathrm{D}=\frac{4 \pi}{3} \rho_\mathrm{g} s^2 v c  \, ,
\end{equation}
where  $\rho_\mathrm{g}$ is the gas density, $s$ is the (spherical) grain
radius, $c$ is the sound speed, and $v$ is the velocity difference between 
dust and gas.
For details of how the equations of motion and the density of
the two fluids are calculated, we refer the reader to BF05.

The dust particles are incompressible ($\rho_\mathrm{d}$ = constant) and 
there is no grain evaporation or coagulation, nor any gas condensation.  All 
simulations presented consider just one (spherical) grain size at a time.


We set up a disk of gas and dust with a total mass of $M_{\rm disk}$ around a 
$1 \msol$ star with an embedded planet of mass $M_{\rm p}$ at a distance of 
$r_{\rm p}$.  The dust phase is 1\% of the total disk mass and the system is 
evolved for about 100 planetary orbits. 
The disk equation of state is isothermal with constant vertical temperature
and radial profile $T \propto r^{-1}$. Code units are set by
$G=M_{\star}=r_{\rm p} = 1$ and the isothermal sound speed at $r=1$ is 
$c = H/r = 0.05$.
The initial density profile is constant and the dust density is $\rho_d$ (see below for details). 
The planet is treated as a point mass particle which moves under the
gravitational influence of the star on a fixed circular orbit  
(i.e. no migration).

Simulations start with 50,000 gas and 50,000 dust particles.
Particles are removed from the simulations if they cross the Hill 
radius of the planet, get closer than 0.4 code units of the central star 
(which sets the inner disk edge), or if they escape past 
4 code units (which sets the outer disk edge).


In this work we present the results of two different disk models: 
(1) a small, low mass disk close to the minimum mass solar nebula 
(MMSN model), and (2) a classical T~Tauri star disk (CTTS model).  
The MMSN disk has a mass of 2.9 10$^{-3}$~$\msol$ and extends from 
2~AU to 20~AU. The standard MMSN model has a 1 Jupiter mass ($M_{\rm Jup}$) 
planet on a circular orbit of radius 5.2~AU in a disk containing grains 1~m 
in size and $\rho_{d}=1.25$~g cm$^{-3}$. 
The CTTS disk mass is 0.01 $\msol$ and spans 16~AU to 150~AU 
in radius.  The standard CTTS model has a $5~M_{\rm Jup}$ planet on a circular 
orbit of radius 40~AU in a disk containing grains 1~mm in size and $\rho_{d}=1.0$ g cm$^{-3}$.  

For both disks we start by running the standard model and compare the
evolution of the gap in the gas and dust phases. We then run a
series of experiments to study the effect of grain size in the dust disk 
for both models, with $s =$ 1~cm, 10~cm and 1~m for the MMSN disk and  
$s=100$~$\mu$m, 1~mm and 1~cm for the CTTS disk.  
(Because the nebula conditions and particularly the density in the 
CTTS and MMSN models are different, 
the grain sizes used in the two models are different in order to 
obtain similar values for the gas drag and hence dust settling 
and migration rates.)
This is followed by a series of experiments that study the effects of 
planetary mass on gap formation and evolution, with 
$M_{\rm p}=0.05, 0.1, 0.2, 0.5$ and 1.0 $M_{\rm Jup}$ for the 
MMSN model and $M_{\rm p} =$0.1, 0.5, 1.0 and 5.0 
$M_{\rm Jup}$ for the CTTS model. 
%

\begin{figure*}[t]
{
\includegraphics[width=8cm]{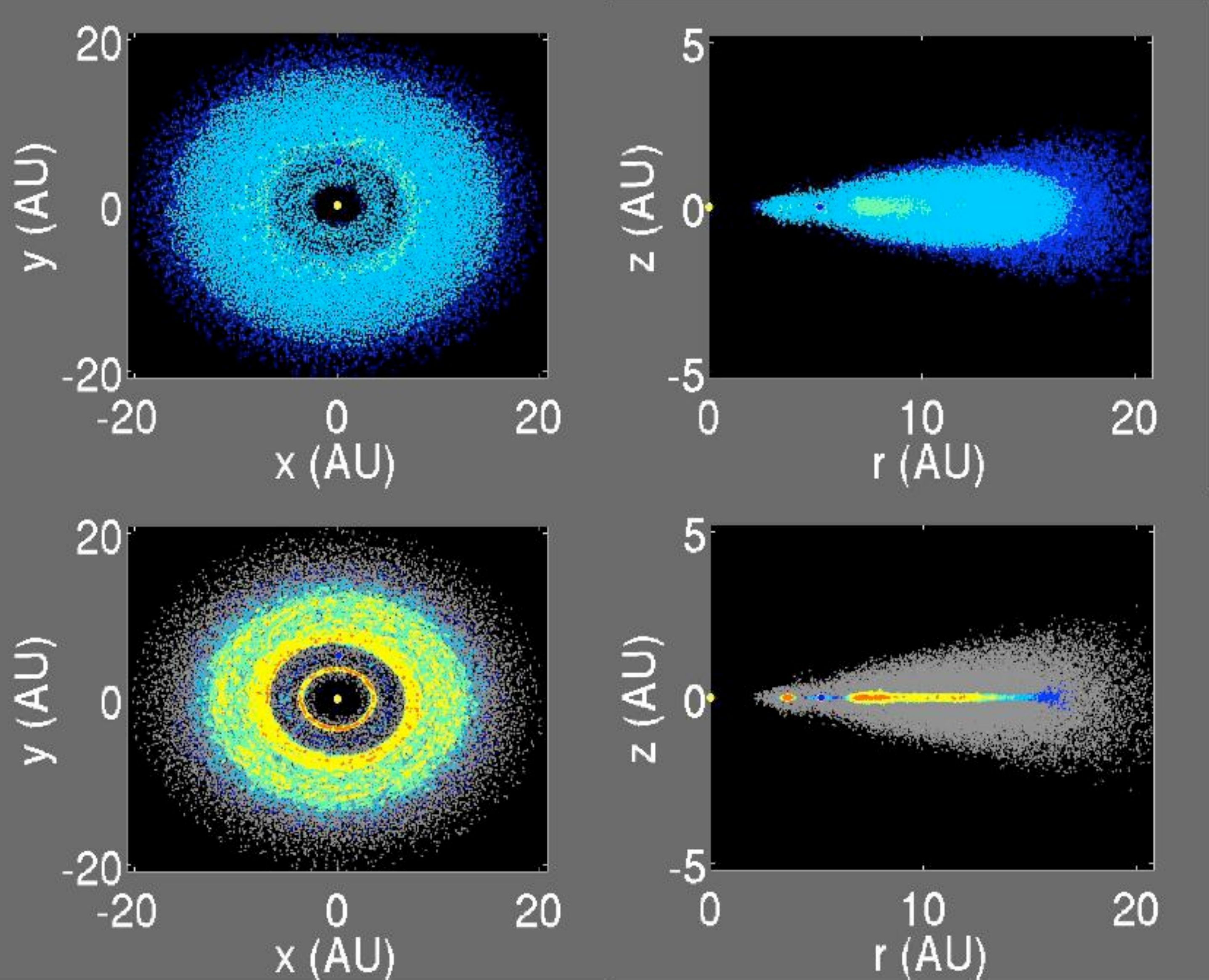}
\includegraphics[width=8cm]{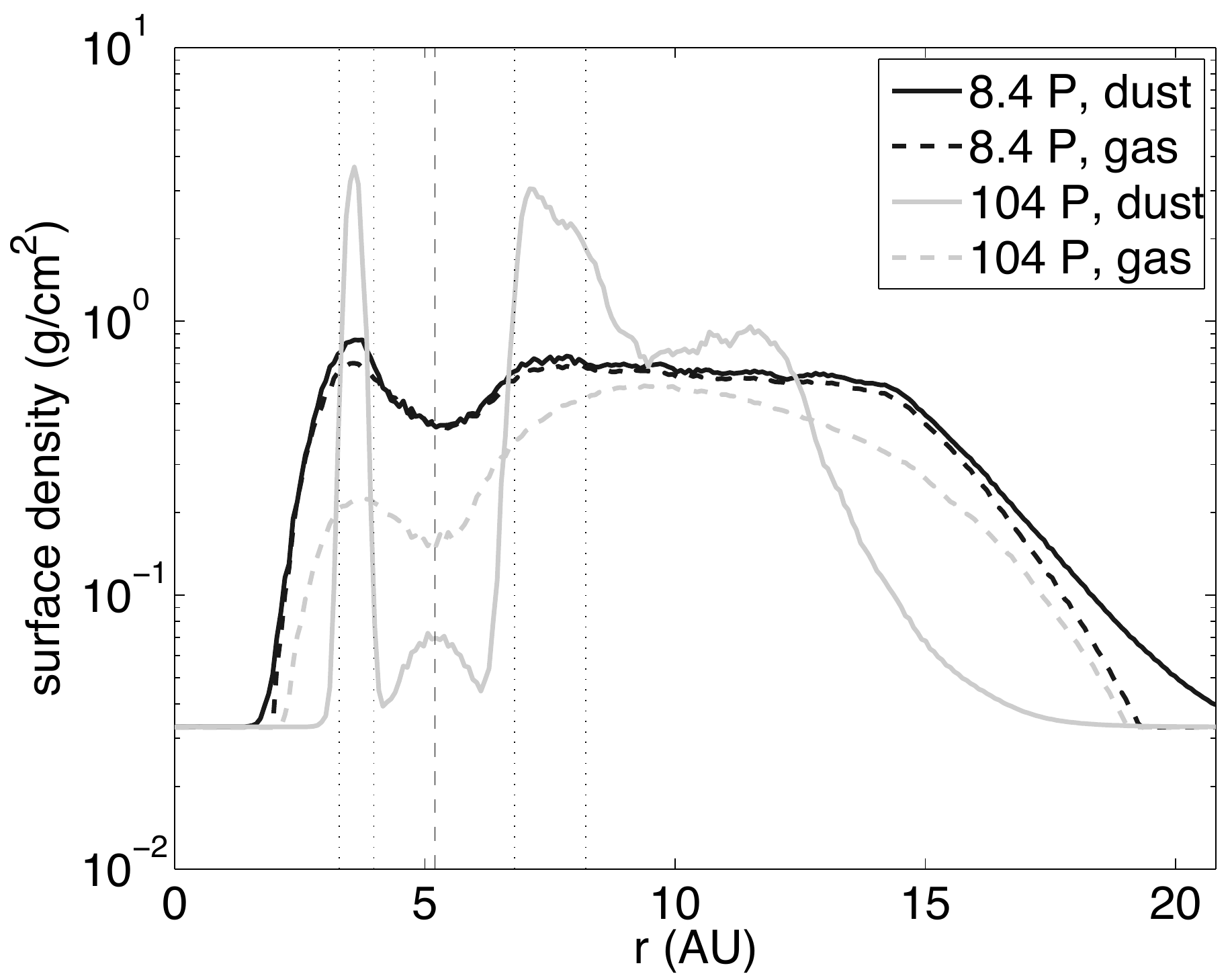}
}
\caption{Left panel: End state of the standard MMSN model after 104 orbits 
(1,230~years), showing top-down and side-on views of the disk. The top row  
shows the gas disk, while the bottom row shows the (coloured) dust overlaid 
on the gas (grey). Right panel: Azimuthally averaged surface density profiles 
for a gap created in the standard MMSN model, showing  the comparison of the 
gas and dust surface densities after 8.4 and 104 orbits. The gas density is
scaled by 0.01 for direct comparison to the dust. The vertical lines 
are, from left to right, the 1$:$2 and 2$:$3 internal resonances, the 1:1 
planetary orbit, and the 3$:$2 and 2$:$1 external resonances respectively.
}
\label{fig-MMSN-standard}
\end{figure*}

\begin{figure*}[t]
{
\includegraphics[width=8cm]{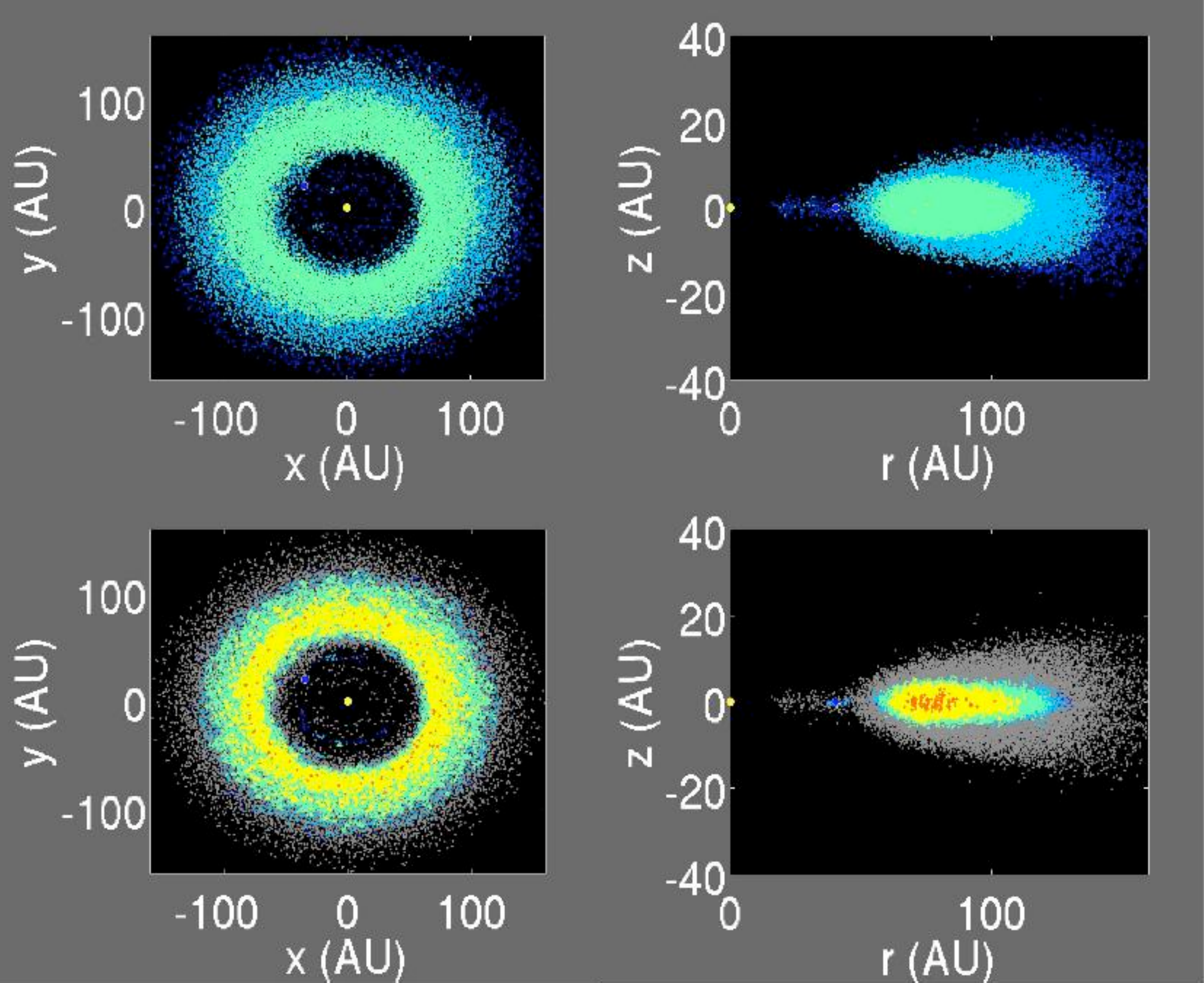}
\includegraphics[width=8cm]{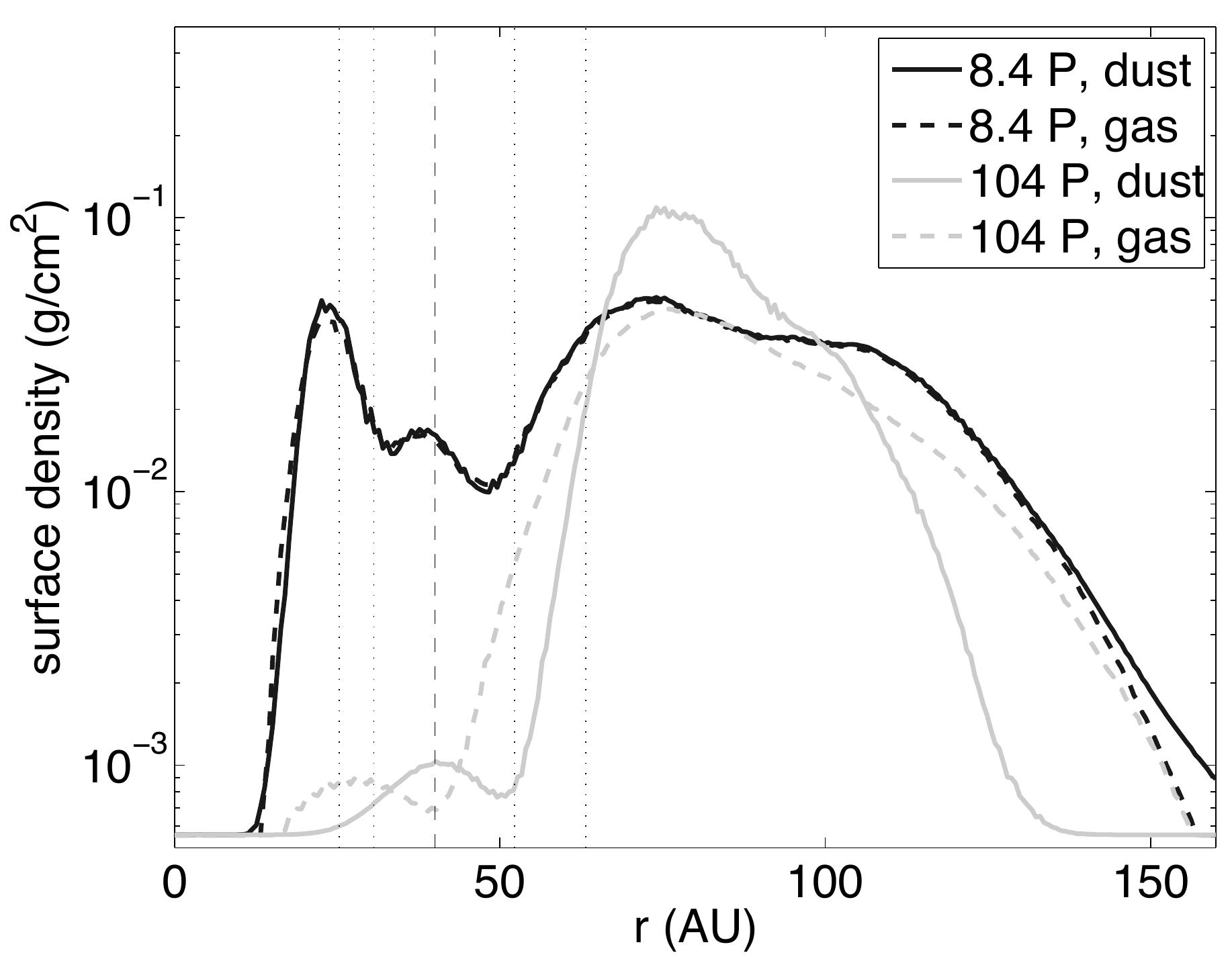}
}
\caption{Same as Fig.\ref{fig-MMSN-standard} but for the standard CTTS model,
  in which 104 orbits is equivalent to 26,310~years. }
\label{fig-CTTS-standard}
\end{figure*}

\section{Simulation results}

The results of the standard MMSN model is shown in
Fig.~\ref{fig-MMSN-standard}.
The left panel shows the top-down ($x,y$) and side-on ($r,z$)
view of the gas and dust disk morphologies. 
While the planet opens a gap in the gas, the gap in the 
dust layer is much more striking.  The right panel of
Fig.~\ref{fig-MMSN-standard} compares the evolution of the azimuthally
averaged surface density profile of the gas and dust after 8.4 and 104 planetary orbits.

%
Recent observations have suggested that massive planets at large distances
from the star may exist, such as a 5~$M_{\rm Jup}$ planet  30~AU from LkCa 15 
\citep{Pietu-etal2006} and a 12.5~$M_{\rm Jup}$ planet 135~AU from 
GG~Tau \citep{Beust05}. 
Our standard CTTS model has a 5~$M_{\rm Jup}$ planet at 40~AU and the 
results are shown in Fig.~\ref{fig-CTTS-standard}.
For the nebular parameters used, such a massive planet almost
completely empties the inner disk of both gas and dust 
\citep[though this is likely due to the large inner disk radius - see][]{CridaMorb2007}.

\subsection{Effect of grain size}
\label{subsec-grainsize}

Since the gap criterion is partially governed by the disk scale height, and the dust
scale height varies with grain size, we ran a series of simulations to determine the 
effect of grain size on the gap. Three grain sizes are 
tested in both disk models: 1~cm, 10~cm and 1~m for the MMSN disk and 
100~$\mu$m, 1~mm and 1~cm for the CTTS disk. 


\begin{figure*}
\includegraphics[width=8cm]{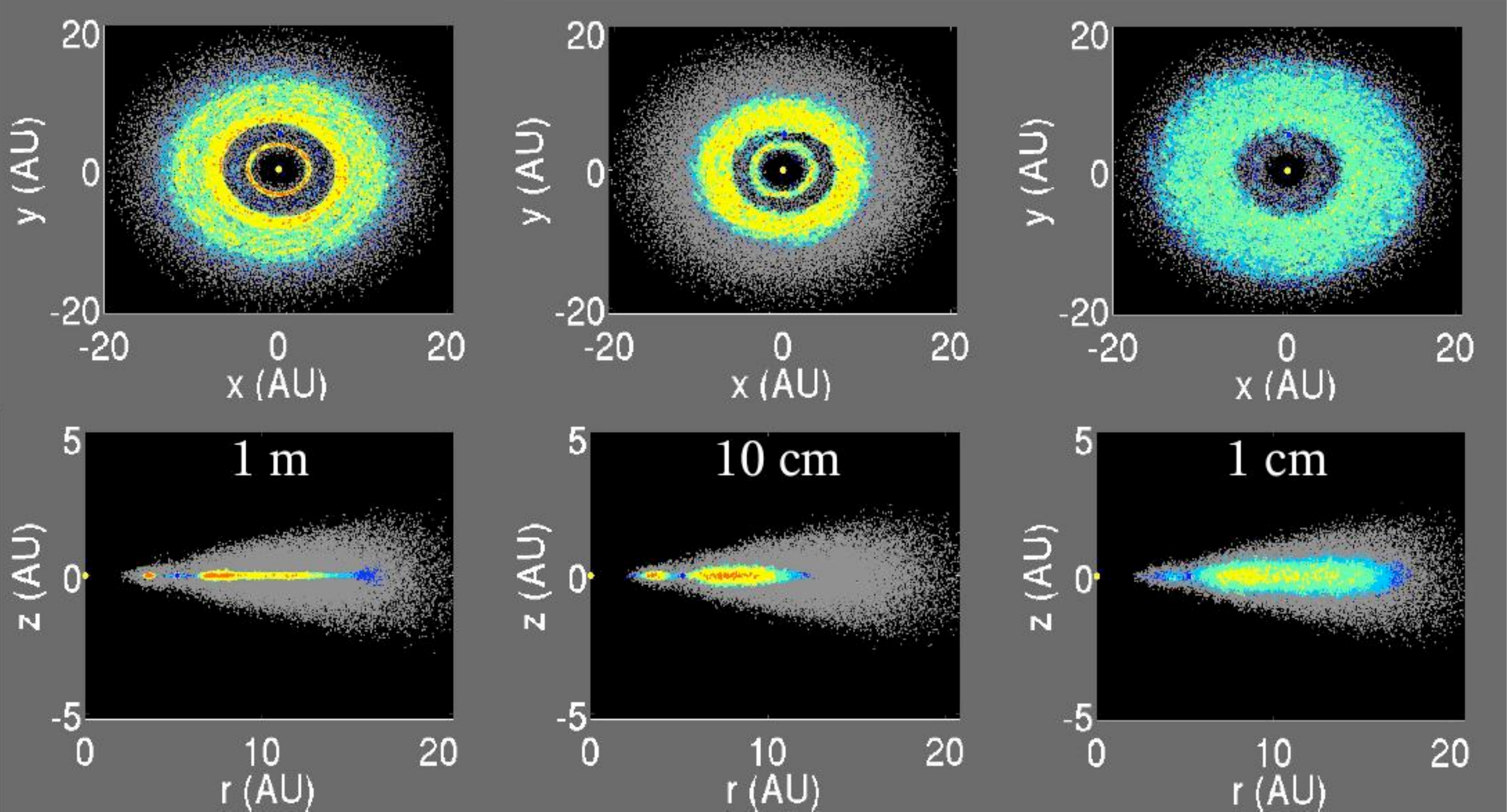}
\includegraphics[width=8.8cm]{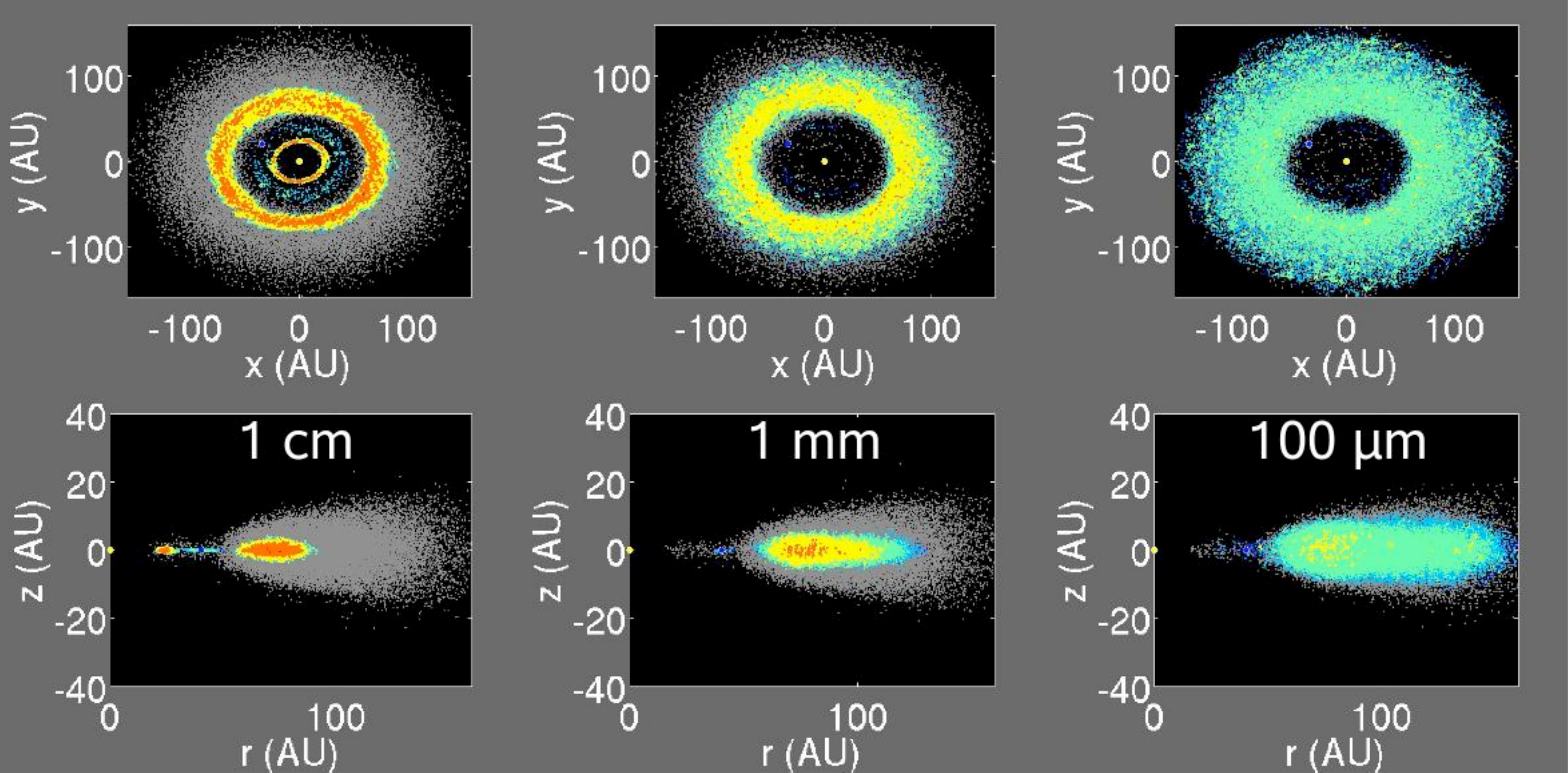}
\caption{Gap created in the MMSN disk (left) and CTTS disk (right) for a range
  of dust grain sizes after 104 orbits.    Top panel shows the top-down view
  of disk and bottom panel shows the side-on view. Grey is gas and the dust is 
  coloured by density. For the MMSN disk (with a 1~$M_{\rm Jup}$ planet at 
  5.2~AU) from left to right shows 1~m, 10~cm and 1~cm sized grains.  For the 
  CTTS disk (with a 5~$M_{\rm Jup}$ planet at 40~AU) from  
  left to right shows 1~cm, 1~mm and 100~$\mu$m sized grains.}
\label{fig-grains3D}
\end{figure*}

Fig.~\ref{fig-grains3D} shows how the gap morphology of both the MMSN and CTTS disks 
vary with dust grains size.
We find that both the width and depth of the gap increases with increasing
grain size.

\subsection{Effect of planetary mass}
\label{subsec-planetmass}

\begin{figure*}[t]
{
\includegraphics[width=7cm]{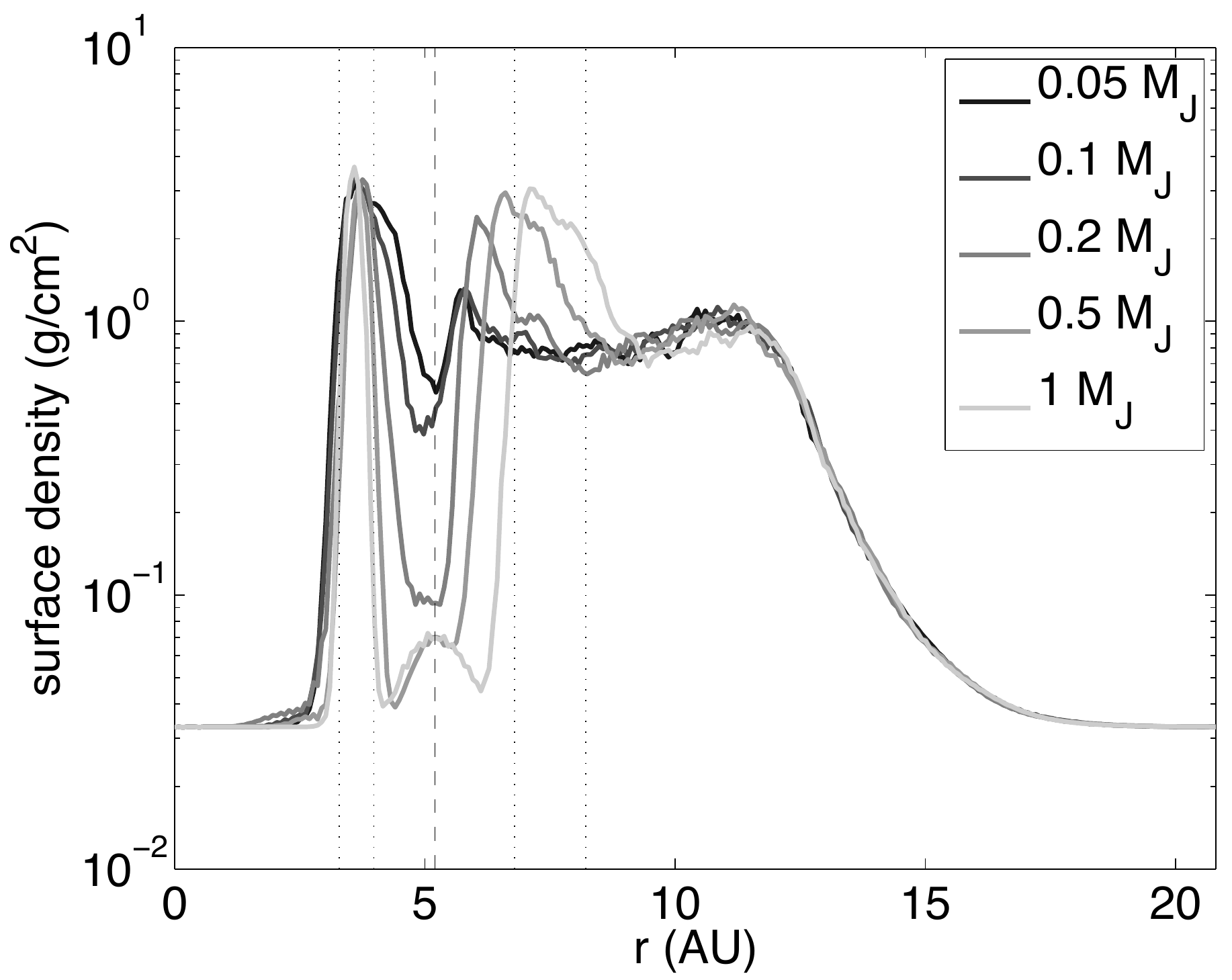}
\includegraphics[width=7cm]{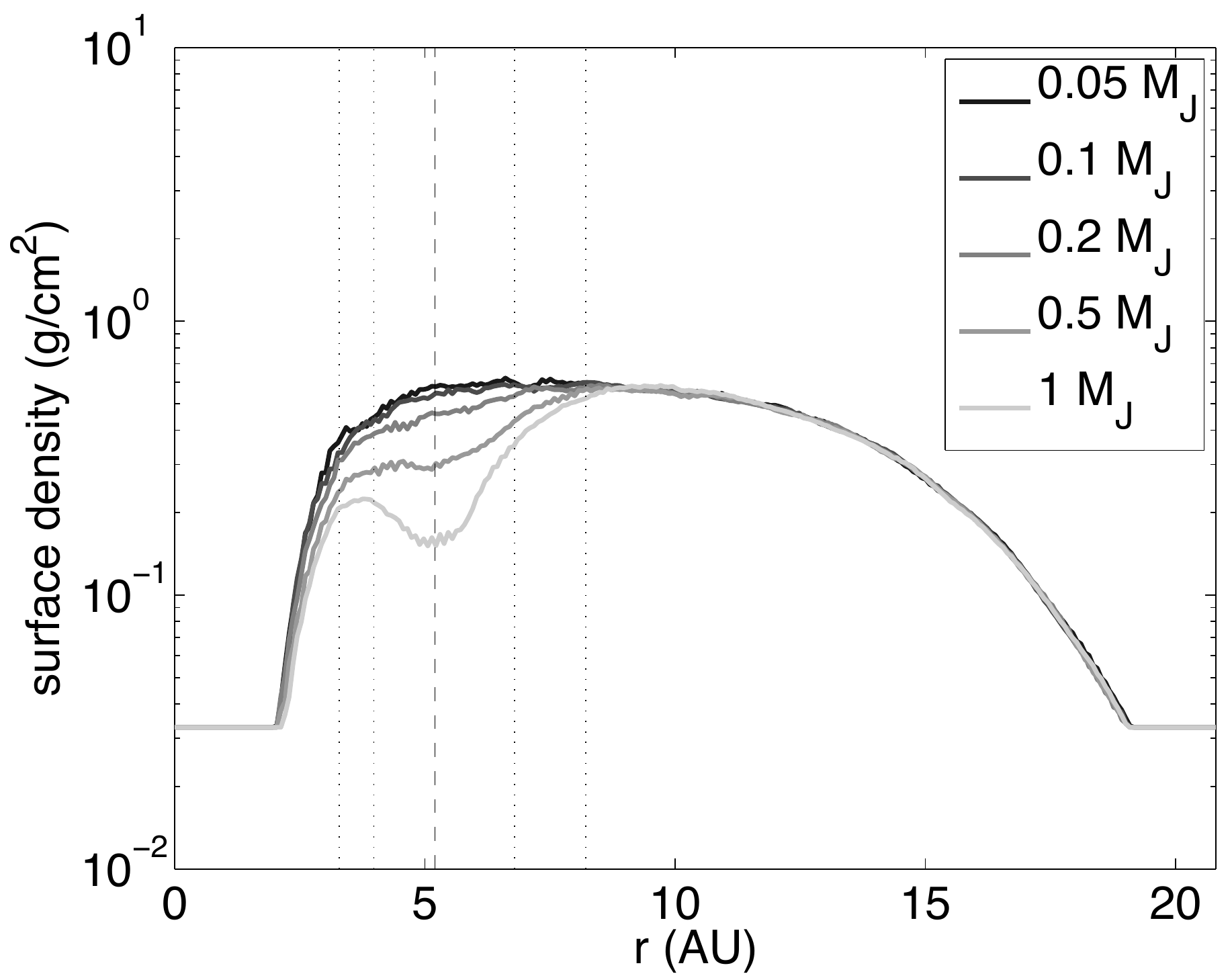}

\includegraphics[width=7cm]{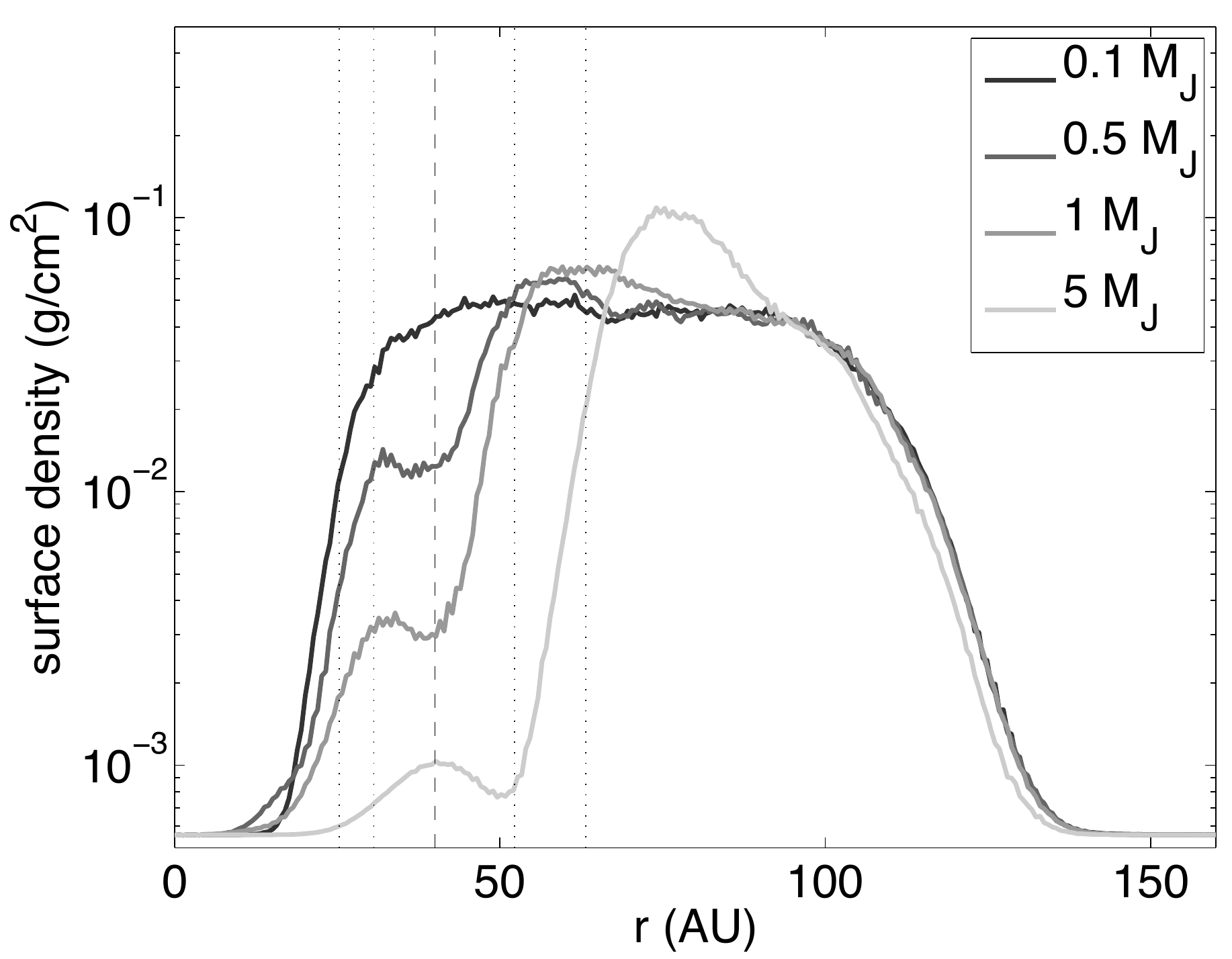}
\includegraphics[width=7cm]{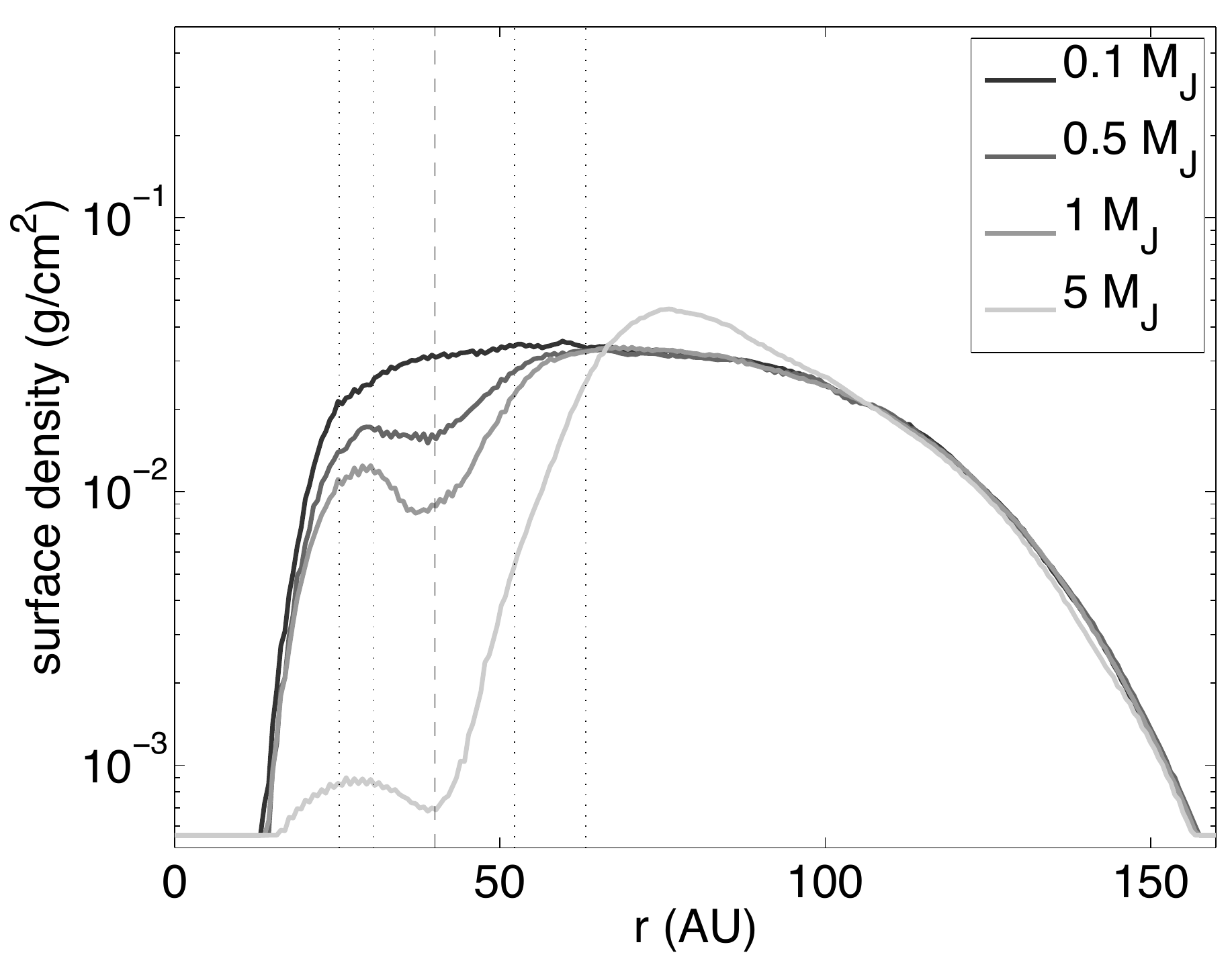}
}
\caption{Azimuthally averaged surface density profiles for various planet
  masses after 104 planetary orbits. Top row: MMSN disk, Bottom row: CTTS 
  disk. Left frames: dust profiles, Right frames: gas profiles.}
\label{fig-planet-2Ddens}
\end{figure*}

Finally we investigate the effect of the planetary mass on the evolution of the
gap.  For our MMSN disk with 1~m grains, $M_p$ varies from
0.05~$M_{\rm Jup}$ to 1~$M_{\rm Jup}$,
and for our CTTS disk containing 1~mm grains,  $M_p$ varies
from 0.1, 0.5, 1 to 5~$M_{\rm Jup}$.  Fig.~\ref{fig-planet-2Ddens} shows the
comparison of the surface density profiles for both models.
For the MMSN disk, the gap is more striking in the dust than the gas, while in
the CTTS disk the inner disk appears depleted of both gas and dust,
while the dust pile-up at the outer gap edge is clearly seen. For the CTTS disk, 
no change is seen in the 
surface density for a 0.1 $M_{\rm Jup}$~planet in either the gas or dust phase.
These results are in general agreement with the minimum planet mass required
to produce a gap in the disk models of \citet{Paardekooper06}.

\section{Discussion and conclusions}

Structures created by planets in dusty disks are more diverse than those
created in the gaseous disks.  With only aerodynamic drag, we find that it is 
possible to create disks with a large central hole or a ring.
\citet{Rice-etal2006} also found that the presence of a planet can produce
disks with a central hole  for certain grain sizes.

Our results have implications for observational predictions of 
protoplanetary disks hosting planets. \citet{Wolf05} and 
\citet{Varniere-etal2006} use results of 2D hydrodynamic simulations 
to produce synthetic images of protoplanetary disks, but these
simulations assume that the gas and dust are well mixed, which our 
results clearly demonstrate is not the case. 

Because our general findings show than the gap is generally more striking 
in the dust disk, we suggest that predictions of observations of
protoplanetary disks are too pessimistic. 
Our results show that the density contrast around the gap can be very 
strong (and the volume density can actually be greater than the gas volume density) 
and this would be detectable with ALMA. Our  
simulations support the results of \citet{Varniere-etal2006}, 
as we clearly see density enhancements in the outer gap edge of 
the CTTS simulations, even for the smaller grains sizes which would 
be responsible for the majority of the sub-mm and mm emission.
Our results also support the predictions of \citet{Paardekooper06}
that gaps created by $0.05 M_{\rm Jup}$ planets in MMSN disks 
should be visible with ALMA.
For a more detailed analysis, we refer the reader to Fouchet et al. (2007).

While we see a clear density increase in the vicinity of the external 3$:$2 resonance 
of our standard MMSN disk, we do not believe that particles are trapped in the
resonance. Plotting the dust eccentricity against semi-major axis when 
drag was neglected clearly shows  resonances 
as thin vertical lines and a V-shaped pattern at the edges of the
gap. However when drag is included, we find that the drag efficiently damps 
high eccentricities and the resonant signatures disappear.
Furthermore, while the dust pile up appears to coincide with the 3$:$2 
external resonance for the MMSN disk (when gas drag is included), this is 
not true for the standard CTTS disk -- see Fig.~\ref{fig-CTTS-standard}.
Thus the accumulation of dust that we and other
authors \citep{Paardekooper06,AlexanderArmitage2007}
notice close to the outer gap edge is not due to resonant trapping.
The accumulation of grains at the external edge of the gap may, 
however, favour the growth of planetesimals in this high density region.

We have conducted a series of 3D numerical simulations of two-phase 
(dust$+$gas) protoplanetary disks to study the behaviour of the dust in the 
presence of a planet. We ran a series of experiments with a minimum mass solar 
nebula disk as well as a larger, more massive Classical T~Tauri star disk, 
varying the grain size and the planet mass.
We find that gap formation is more rapid and striking in the dust layer than 
in the gas layer. Varying the grain size alone results in a variety of 
different structures, and for the CTTS disk these differences will be 
detectable with ALMA. For low mass planets in our MMSN disk, a gap 
was found to open in the dust layer while not in the gas layer. 
Simulations like these 
can be used to help interpret observations to constraint the planet mass 
and grain sizes in protoplanetary disks.



\end{document}